\pdfoutput=1
\documentclass[a4paper,11pt]{article}
\usepackage{amsmath,amssymb}
\usepackage{pos}
\usepackage{hhline}
\usepackage{tabularx}
\usepackage{hyperref}
\newcommand{\mumut}{\ensuremath{\mu^+\mu^-\to}}

\title{Precision test of the muon-Higgs coupling at a high-energy muon collider}
\ShortTitle{The Muon-Higgs coupling at a muon collider}

\author*[a]{J\"urgen Reuter}

\affiliation[a]{Deutsches Elektronen-Synchrotron DESY, Notkestr. 85,
  22607 Hamburg, Germany.}

\emailAdd{juergen.reuter@desy.de}

\abstract{
  Muon colliders offer the possibility to go to very high energies
  with relatively small circular colliders, energies up to 10 or 14
  TeV are envisioned. Due to their very clean collider environment
  they provide a fantastic tool to search for new physics in the
  electroweak sector, especially through the production of multiple EW
  vector and Higgs bosons, and they allow to measure the Higgs-muon
  coupling very precisely. I will elucidate the physics capabilities
  from these processes and also discuss issues on precision
  predictions for SM backgrounds at high-energy lepton colliders.
}

\FullConference{%
  Corfu Summer Institute 2022 "School and Workshops on Elementary Particle Physics and Gravity",\\
  28 August - 1 October, 2022\\
  Corfu, Greece}

\begin{document}
\maketitle

\section{Introduction}

Muon colliders have recently gained a lot of interest again driven by
technological progress made on the accelerator side. The muon collider
is now not any longer considered as a Higgs factory operating at the
Higgs threshold, but at an energy frontier machine in the multi-TeV
range which very prominently featured in the US Snowmass Community
Summer Study~\cite{MuonCollider:2022nsa,MuonCollider:2022xlm,
  MuonCollider:2022glg,Accettura:2023ked}. 
Due to the very clean environment of the pointlike leptonic beams,
their search reach for new particles and phenomena can supersede that
of high-energy hadron colliders like the FCC-hh (one of these examples
is the search for heavy neutrinos, cf.~\cite{Mekala:2023diu}). Muon
colliders do have much less synchrotron radiation and much less
beam-energy spread than high-energy (i.e. linear) $e^+e^-$ colliders,
as well as much less QED initial-state radiationm which leads to
different physics reaches between MuC and e.g. CLIC at the same
collider energies (cf. e.g.~\cite{Mekala:2023diu}. Here, we focus on
the search for new physics in deviations of the muon Yukawa coupling
from its Standard Model (SM) value. This proceedings article contains
two sections, Sec.~\ref{sec:muyuk} on the sensitivity of muon
colliders to anomalous muon-Higgs couplings and the discovery
potential for higher-dimensional operators in this sector from
multi-boson signatures at the MuC, while in Sec.~\ref{sec:muprec} we
show that it is possible to provide precision predictions for these
signatures in the SM regarding complete electroweak next-to-leading
order (NLO) corrections. Finally, we briefly conclude.

\section{The muon Yukawa coupling sensitivity}
\label{sec:muyuk}

After the H(125) particle discovery in 2012, many of its properties
have been proven to be SM-like in the past decade,
however, the couplings to the second-generation fermions are still
very elusive. In the SM, the muon Yukawa coupling is one of the
smallest parameters. It is expected, that after the high-luminosity
run of the Large Hadron Collider (LHC) this coupling could be measured
with a precision of ca. 5 \%, however, its sign cannot be determined
as it is only accessible from the Higgs decay. A model-independent
measurement would be highly desirable, and indeed it is possible in
direct production at a muon collider. Here, the muon Yukawa coupling
shows a subtle cancellation with production of longitudinal gauge bosons
at high energies. As the muon Yukawa coupling is a tiny parameter in
the SM, huge deviations beyond the SM are possible, and could deceit a
default power counting in SM Effective Field Theory (SMEFT). There are
also BSM examples, like extra-dimensional theories, where the
muon-Yukawa coupling could follow a completely different
renormalization-group running as in the SM, and hence has completely
different values at very energies. Here, we investigate the effect of
non-standard muon-Higgs couplings on multi-boson final states: here,
longitudinal polarizations dominate at high energies and anlytic
calculations can be approximated bu the Goldstone-boson equivalence
theorem. We will parameterize new physics effects by EFT operator
insertions.

We considered two different EFT representations, a non-linear
realization with non-linear sigma model of Goldstone bosons and a
\begin{figure}
  \includegraphics[width=.49\textwidth]{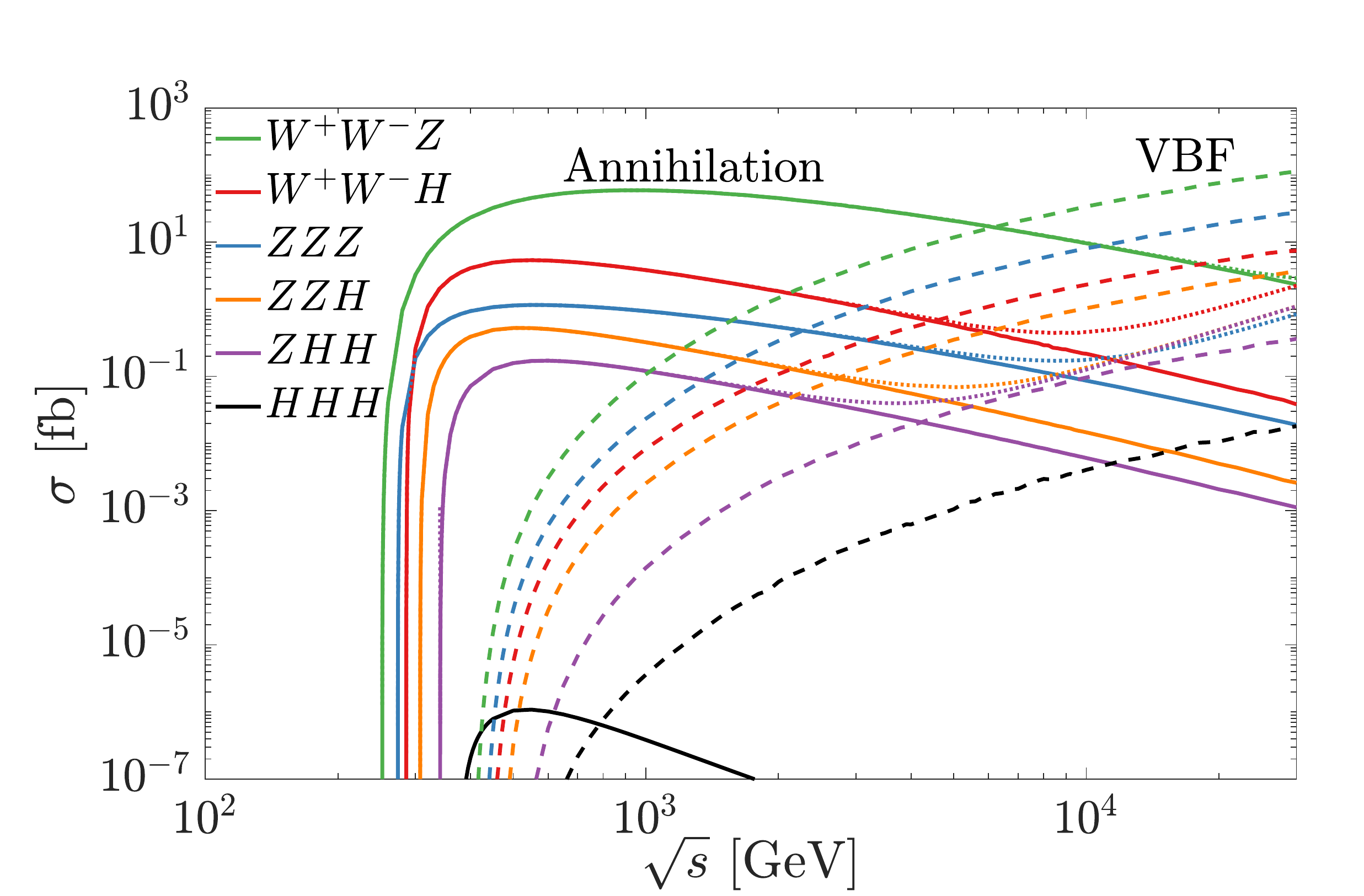}
  \includegraphics[width=.49\textwidth]{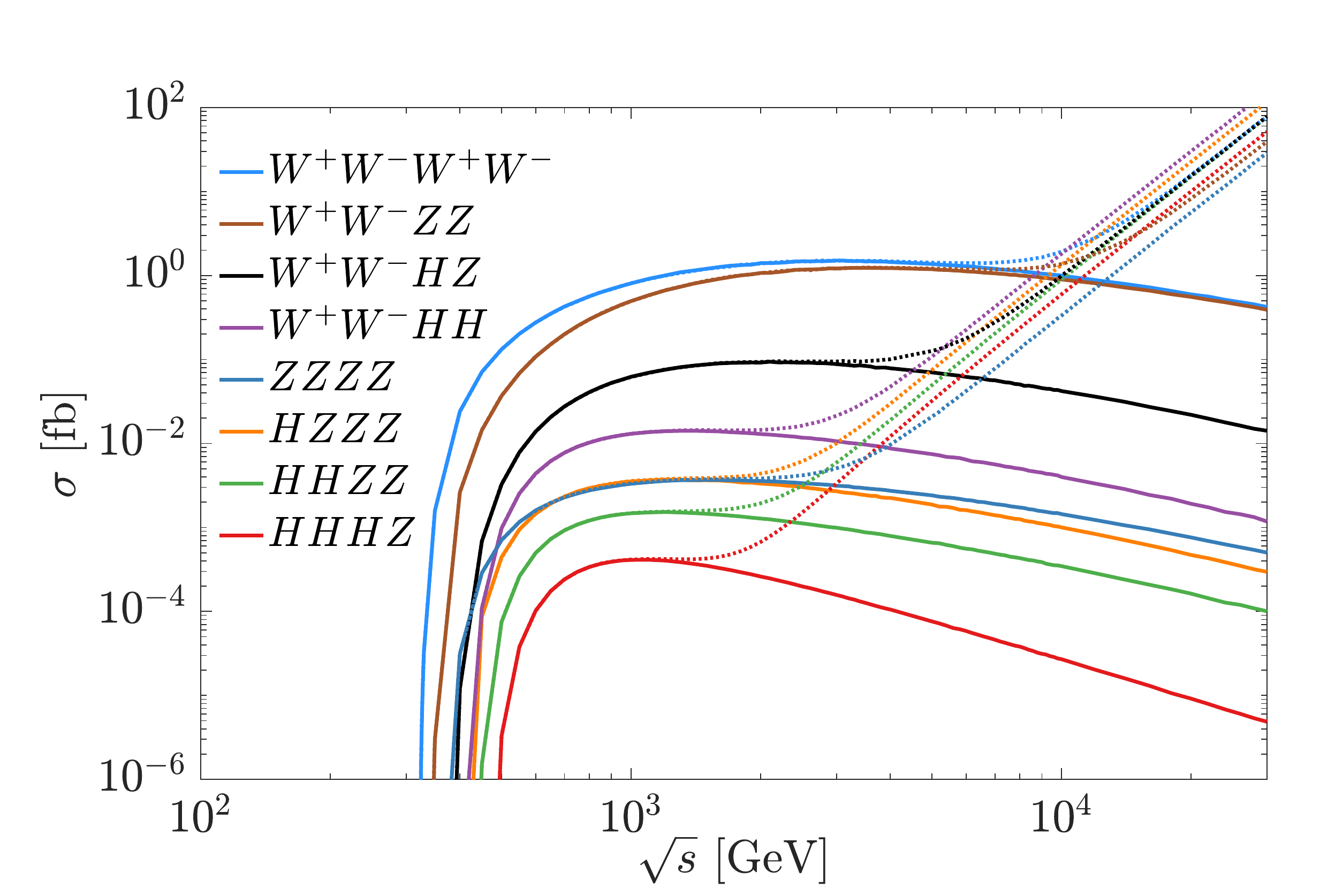}
  \caption{Cross sections of the production of three and four bosons
    as a function of the MuC energy. Full lines are the SM, dotted
    lines show the increase due to higher-dimensional operators, and
    dashed lines cross section from the VBF channel.}
  \label{fig:3_and_4_prod}
\end{figure}
linear representation in terms of a SMEFT expansion. As an extreme
case we assume a toy model where the higher-dimensional operators
completely cancel the SM contribution, leading to a vanishing
muon-Higgs coupling, called "matched" case. The modification of the
muon-Higgs coupling gets the following modifications from its SM
value:
\begin{equation}
  K_\ell =1-\frac{v}{\sqrt 2} M_\ell^{-1}\sum_{n=1}^{\infty}
  C^{(n)}_{\ell\varphi} \frac{n v^{2n}}{2^{n-1}} \quad , 
\end{equation}
with contributions of dimensionality $4 + 2n$. Here, $M_\ell$ is the
muon mass, $v$ the Higgs vacuum expectation value and
$C_{\ell\phi}^{(n)}$ are the Wilson coefficients of the corresponding
higher-dimensional operators. For more details on the
model setup cf.~\cite{Han:2021lnp}.

Among the three-boson processes on which we concentrate here, the
process $\mumut ZZH$ shows the largest deviations from the SM cross
section, however, has a smaller rate than the process $\mumut
W^+W^-H$. For this reason, we mainly use the latter process in this
study.

In Fig.~\ref{fig:3_and_4_prod} we show the cross sections for the
production of three and four gauge or Higgs bosons for the SM (full
lines, promt production), dashed lines (SM, vector-boson fusion, VBF)
and dashed lines with insertion of higher-dimension operators as a
function of the MuC energy. The VBF channel is not sensitive to the
(anomalous) muon Yukawa coupling and needs to be filtered out in the
event selection. This is achieved with a cut-based selection,
demanding that the total invariant mass of the 3- or 4-boson systems
\begin{figure}
  \includegraphics[width=.49\textwidth]{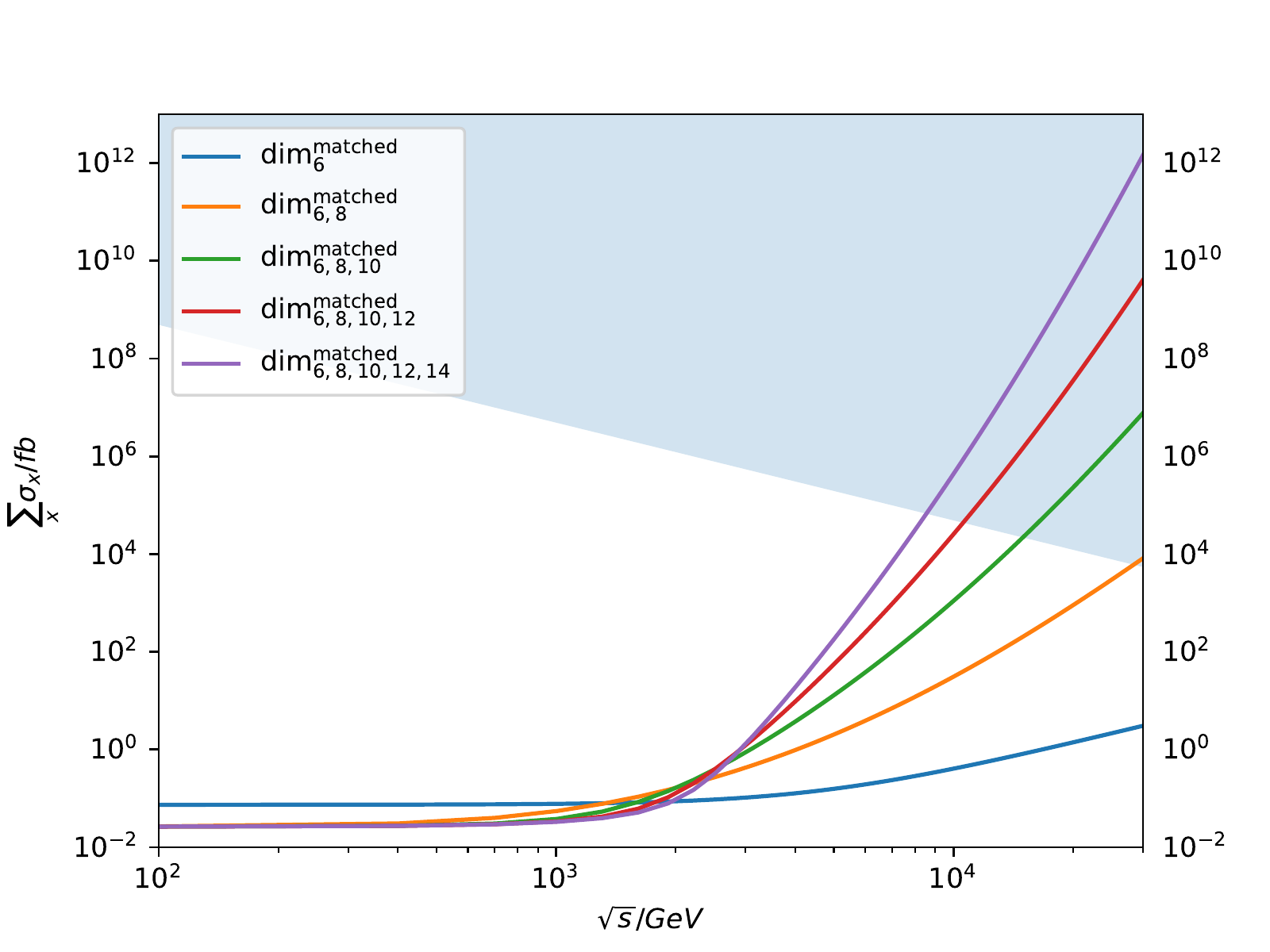}
  \includegraphics[width=.49\textwidth]{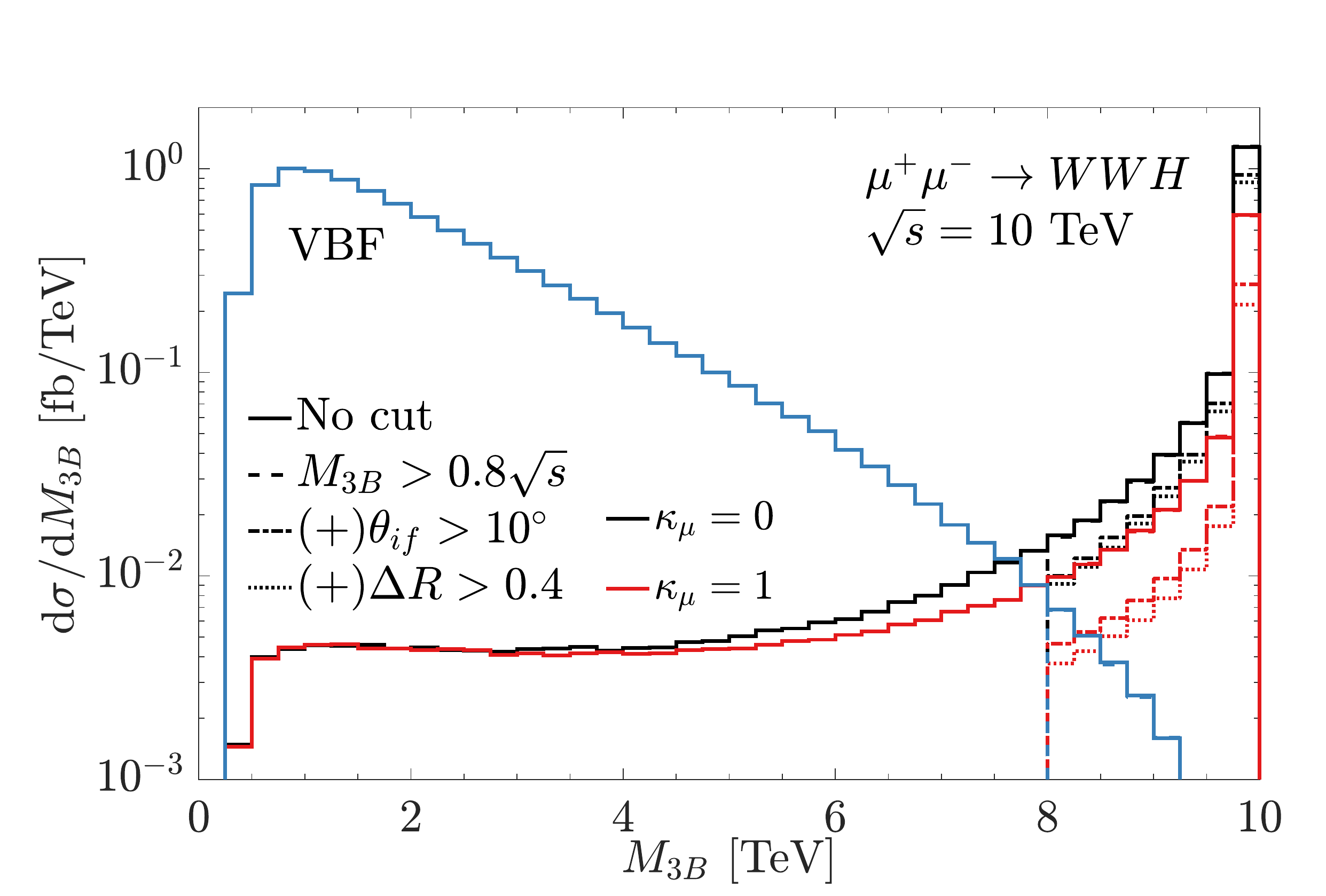}
  \caption{Left plot: Unitarity bound for the matched cases
    considering Wilson coefficients from different combinations of
    higher-dimensional operators as a function of the MuC
    energy. Right plot: differential distribution of the invariant
    mass of the 3-boson system in $\mumut W^+W^-H$ at 10 TeV: prompt
    production vs. VBF, and the effect of the fiducial phase-space
    cuts.}
  \label{fig:unitarity_m3b}
\end{figure}
exceeds 80\% of the total energy, $M_{3/4B} > 0.8\cdot\sqrt{s}$,
requesting a minimum polar angle of 10 degrees, $\theta_{if} >
10^\circ$ for each boson (as the VBF process originates from collinear
splitting being mostly forward), and a minimal  $R$-space distance, $\Delta
R_{BB} > 0.4$ as resolution criterion. The separation of signal and
VBF background is shown in the right hand side of
Fig.~\ref{fig:unitarity_m3b} for the invariant mass of 3-boson system
for the process $\mumut W^+W^-H$, where one sees that the first cut
above is really efficient.

There are bounds from perturbative partial wave unitarity for the
processes of several Higgs bosons on the size of operator coefficients
or total cross sections. These are shown in the left plot of
Fig.~\ref{fig:unitarity_m3b} for the matched cases (i.e. vanishing
explicit muon-Higgs coupling) for the case of dimension-6 operators
only, and then including dimension-8, -10, -12 and -14 operators,
respectively. For the most interesting cases of dimension-6 and -8,
the considered MuC energies of $1 < \sqrt{s} < 30$ TeV are still
within the bounds.

Our results have been cross-checked with two independent analytic
calculations in the Goldstone-boson equivalence approximation as well
as by numeric simulation, using the Monte-Carlo event generator
WHIZARD~\cite{Kilian:2007gr,Moretti:2001zz}, using a UFO model
implementation~\cite{Degrande:2011ua,Darme:2023jdn}. Several different
benchmerk scenarios have been studied: dimension-6 alone, dimension-8
alone, dimension-6+8 combined as well as the abovementioned "matched"
case. MuC energies between 1 and 30 TeV have been considered with the
luminosity scaling of $\mathcal{L} = \left( \sqrt{s} / 10\,\text{TeV}
\right)^2 \cdot 10 \text{ab}^{-1}$. The numer of signal events is
defined by the number of events with signal-strength modifier
$\kappa_\mu$ subtracted by the number of corresponding SM events, $S =
N_{\kappa_\mu} - N_{\kappa_\mu = 1}$. The number of background events
is given by $B  = N_{\kappa_\mu = 1} + N_{\text{VBF}}$ is this SM
event number and the number of VBF events in the fiducial phase space
volume defined above. Our significance is then defined via
$\mathcal{S} = S/\sqrt{B}$, where the number of events in the SM is
always the smallest: $N_{\kappa_\mu} \geq N_{\kappa_\mu = 1}$. Also
note that because the deviations of the cross sections is the same for
up- and downwards deviations from the SM value, $\kappa_\mu = 1 \pm
\delta$, the significance is the same for these two value. The 5$\sigma$
significance for establishing such a deviation from a measurement of
\begin{figure}
  \centering
    \includegraphics[width=.6\textwidth]{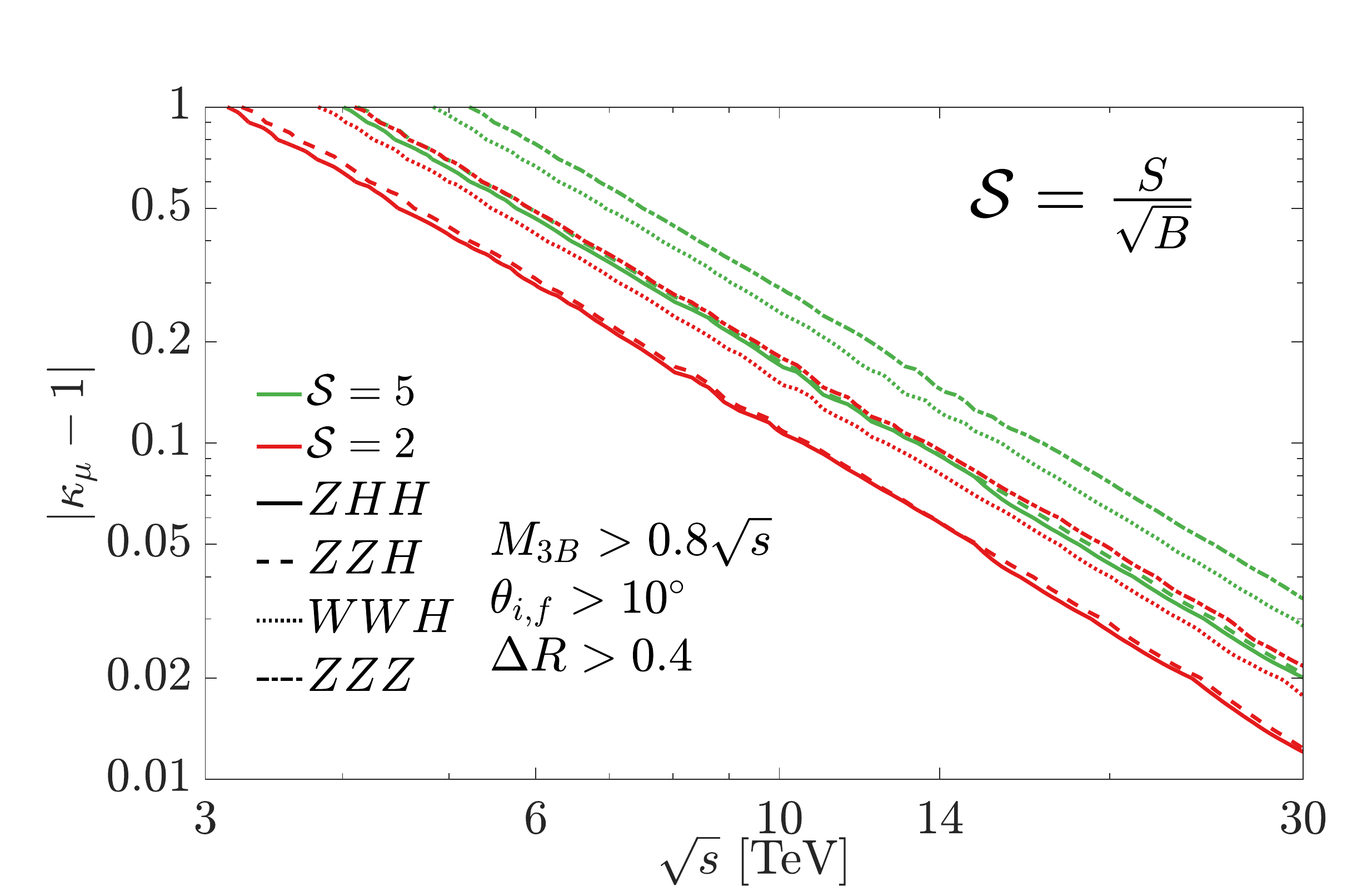}
  \caption{2$\sigma$ and 5$\sigma$ significance reach for different
    deviations of the muon-Higgs coupling from its SM value as a
    function of the muon collider energy. Four different triple-boson
    processes are shown, within the same fiducial phase-space volume.}
  \label{fig:significance_reach}
\end{figure}
these processes is roughly possible for a 20\% deviation at a 10 TeV
MuC, and goes down to a 2\% deviation at a 30 TeV MuC. This translates
to an access to new physics scales of several tens of TeV to almost a
hundred TeV:
\begin{equation}
  \Lambda \; > \; 10 \, \text{TeV}
  \,\cdot\,\sqrt{\frac{g}{\Delta\kappa_\mu}} \qquad.
\end{equation}
This reach is summarized in Fig.~\ref{fig:significance_reach}.

\section{Precision predictions for multi-boson processes at the muon
  collider}
\label{sec:muprec}

To match the experimental precision at colliders, high-precision
higher-order theoretical predictions and simulations are
indispensable. In the last section, we discussed multi-boson
signatures, $\mu^+\mu^- \to V^n H^m$ with $V \in \{W^\pm,Z\}$
and $n+m \leq 4$, as discovery probes for new physics in the
electroweak sector. Electroweak processes at high energies experience
large electroweak corrections in terms of Sudakov logarithms (double
and single logarithms of the form $\log \mu^2/M_V^2$) that are
typically negative and reduce the cross section, especially in the
high-$p_T$ tails of final-state particles. These Sudakov logarithms
originate from two different sources: (1) virtual electroweak
corrections are not cancelled fully with corresponding real
corrections, as one only consider real QED radiation, but no $W/Z$
radiation, and (2) the initial state, $\mu^+\mu^-$, and hence also the
final state is not an electroweak singlet. Leading EW corrections can
be approximated by these Sudakov logarithms (and they also provide
means of their resummation), but a complete fixed-order NLO EW
calculation is nevertheless necessary to provide high precision in
all regions of phase space, and not only the kinematically enhanced
ones. In Ref.~\cite{Bredt:2022dmm}, all relevant processes with two,
three and four EW gauge or Higgs bosons have been computed. For
brevity, we show here the result only for 3 TeV center-of-mass energy,
with the result given in Tab.~\ref{tab:ew_nlo}. For the 10 TeV results
\begin{table}
  \centering
  \small
  \begin{tabularx}{0.8\textwidth}{l|r|r|r}
    $\mumut X, \sqrt{s}=3$ TeV  &
    $\sigma_{\text{LO}}^{\text{incl}}$ [fb]  &
    $\sigma_{\text{NLO}}^{\text{incl}}$ [fb] &
    $\delta_{\text{EW}} $ [\%]\\
    &&&
    \\\hline\hline
    $W^+W^-$       &        $  4.6591(2)\cdot 10^2 $      &   $ 4.847(7) \cdot 10^2 $    & $ +4.0(2) $ \\
    $ZZ$      &       $ 2.5988(1) \cdot 10^1$      &  $ 2.656(2) \cdot 10^1 $    & $ +2.19(6) $ \\
    $HZ$   &     $ 1.3719(1) \cdot 10^0 $   &   $ 1.3512(5) \cdot 10^0 $ & $  -1.51(4)$ 
    \\\hline

    $ W^+W^-Z$  &   $ 3.330(2) \cdot 10^1 $       &     $ 2.568(8) \cdot 10^1 $   &   $ -22.9(2) $ \\
    $W^+W^-H$      &     $ 1.1253(5) \cdot 10^0 $    &      $ 0.895(2)\cdot 10^{0} $   &  $ -20.5(2) $ \\
    $ZZZ$       &   $ 3.598(2) \cdot 10^{-1} $     &  $ 2.68(1) \cdot 10^{-1} $   &  $ -25.5(3)$\\
    $HZZ$      &    $  8.199(4)\cdot 10^{-2} $  &   $ 6.60(3)\cdot 10^{-2} $    & { $ -19.6(3) $} \\
    $ HHZ$   & $ 3.277(1) \cdot 10^{-2} $ &  $ 2.451(5) \cdot 10^{-2} $ & $ -25.2(1)$\\
    $ HHH$     &   $ 2.9699(6) \cdot
    10^{-8} $  &   $ 0.86(7) \cdot 10^{-8}~^* $  &  $  $
    \\\hline

    $W^+W^-W^+W^-$   &   $ 1.484(1) \cdot 10^0 $  &    $  0.993(6)\cdot 10^0 $  &   $ -33.1(4) $ \\
    $W^+W^-ZZ$     &   $  1.209(1)\cdot 10^0 $  & $ 0.699(7) \cdot 10^0 $ &  $ -42.2(6) $ \\
    $W^+W^-HZ$      &       $ 8.754(8)  \cdot 10^{-2} $ & $ 6.05(4) \cdot 10^{-2} $ & $ -30.9(5) $\\
    $W^+W^-HH$    &   $ 1.058(1) \cdot 10^{-2} $      &   $ 0.655(5)\cdot 10^{-2} $   &  $ -38.1(4) $\\
    $ZZZZ$     &   $ 3.114(2) \cdot 10^{-3} $    &     $ 1.799(7)\cdot 10^{-3} $   &  $ -42.2(2) $\\
    $HZZZ$     &   $ 2.693(2)\cdot 10^{-3} $    &     $ 1.766(6)\cdot 10^{-3} $   &  $ -34.4(2) $\\
    {$HHZZ$}     &   $ 9.828(7) \cdot 10^{-4} $    &     $ 6.24(2)  \cdot 10^{-4} $   & {$ -36.5(2)$}\\
    $HHHZ$     &   $ 1.568(1) \cdot 10^{-4} $   &      $ 1.165(4) \cdot 10^{-4} $   &  $ -25.7(2) $\\
  \end{tabularx}
  \caption[gan:ew_nlo]{Total inclusive cross sections
    at LO and NLO EW with corresponding relative corrections
    $\delta_{\text{EW}}$, for two-, three- and four-boson
    production at $\sqrt{s}= 3$ TeV.}
  \label{tab:ew_nlo}
\end{table}
and more technical details confer the original
paper~\cite{Bredt:2022dmm}. The NLO EW results have been calculated
with massive muons, without a collinearly factorized lepton PDF in the
initial state. The calculations rely on the NLO QCD+EW
automation in the Monte Carlo framework WHIZARD~\cite{Kilian:2007gr}
with virtual matrix elements from RECOLA~\cite{Denner:2017wsf}, using
WHIZARD's parallelized phase-space
integrator~\cite{Brass:2018xbv}. This built upon an early EW NLO
implementation for SUSY particles~\cite{Kilian:2006cj,Robens:2008sa}
and NLO QCD applications for high-energy lepton
colliders~\cite{ChokoufeNejad:2016qux,Bach:2017ggt}.
The results have been validated with results from the literature
for electron-positron processes, and the NLO corrections have been
compared to the leading Sudakov corrections and have been found to be
in good agreement. 


\section{Conclusions}

In this proceedings article we have shown the importance of future
high-energy muon colliders for the search of deviations of the
muon-Higgs coupling from its SM value. For collider energies beyond 10
TeV, 5$\sigma$ significance can be reached for deviations at the level
of signal-digit per cent. Unlike the LHC, where this is accessible
only from the decay, the sign of the deviation can be accesses. This
search potential is reach in processes of one to three electroweak
gauge bosons in association with the 125 GeV Higgs boson. Different
processes, e.g. $WWH$ and $ZZH$ final state, show different event
numbers and different signifcance to such deviations. Systematic
uncertainties can be reduced by taking ratios of these processes.

In the second part, we have provided the electroweak one-loop
corrections to the SM processes, which are negative and between 10-40
per cent of the LO cross section: this can be explained with the
importance of electroweak Sudakov logarithms in the high-energy
regime.

\subsection*{Acknowledgments}
\noindent
JRR likes to thank the organizers for a great workshop at the very
pleasant location of the Corfu Summer Institute.
Support is gratefully acknowledged by the Deutsche
Forschungsgemeinschaft (DFG, German Research Association) under
Germamny's Excellence Strategy-EXC 2121 ``Quantum Universe"-39083330.

\bibliographystyle{JHEP}
\bibliography{refs}

\end{document}